\documentclass[aps,prd,12pt,superscriptaddress,eqsecnum,amsmath,nofootinbib,preprintnumbers]{revtex4-2}%

\usepackage{color,graphicx,float,subfigure}
\usepackage{amsfonts,amssymb,theorem,mathrsfs,times}
\usepackage{bm}
\usepackage{mathtools}
\usepackage{amsfonts,amssymb,theorem,mathrsfs}
\usepackage{dsfont}
\usepackage{setspace}
\usepackage{graphicx}
\usepackage{ulem}
\usepackage[colorlinks,linkcolor=blue,anchorcolor=blue,citecolor=blue]{hyperref}
\usepackage{orcidlink}
\usepackage{comment}
\usepackage{subcaption}
\usepackage{caption}  
\usepackage{soul}

\soulregister\cite7
\soulregister\ref7
\soulregister\pageref7

\DeclareRobustCommand{\stmath}[1]{\mbox{\ensuremath{#1}}}
\soulregister\stmath7

{\theorembodyfont{\upshape}
}
{\theorembodyfont{\upshape}
}
{\theorembodyfont{\upshape}
}
{\theorembodyfont{\upshape}
}
{\theorembodyfont{\upshape}
}
{\theorembodyfont{\upshape}
}

\newcommand{\dalm}{\kern1pt\vbox{\hrule height 0.9pt\hbox{\vrule width
0.9pt\hskip 2.5pt\vbox{\vskip 5.5pt}\hskip 3pt\vrule width
0.3pt}\hrule height 0.3pt}\kern1pt}

\begin{document}

\title{Signatures of the Israel Junction II: Double Photon Rings in Slowly Rotating Kerr Spacetime  with Thin Shell}

\author{Long-Yue Li  \orcidlink{0000-0002-7785-1116}}
\email{lilongyue@nbu.edu.cn}
\affiliation{Institute of Fundamental Physics and Quantum Technology, Department of Physics, School of Physical Science and Technology, Ningbo University, Ningbo, Zhejiang 315211, China}

\author{ Li-Ming Cao  \orcidlink{0000-0001-6633-6613}}
\email{caolm@ustc.edu.cn}
\affiliation{
Interdisciplinary Center for Theoretical Study and Department of Modern Physics,\\
University of Science and Technology of China, Hefei, Anhui 230026,
China}

\author{\\Yungui Gong \orcidlink{0000-0001-5065-2259}}
\email{gongyungui@nbu.edu.cn }
\affiliation{Institute of Fundamental Physics and Quantum Technology, Department of Physics, School of Physical Science and Technology, Ningbo University, Ningbo, Zhejiang 315211, China}

\author{ Xia-Yuan Liu  \orcidlink{0009-0008-3852-0147}}
\email{liuxiayuan@mail.ustc.edu.cn}
\affiliation{
Interdisciplinary Center for Theoretical Study and Department of Modern Physics,\\
University of Science and Technology of China, Hefei, Anhui 230026,
China}

\author{Wenting Zhou  \orcidlink{0000-0003-4046-753X}}
\email{zhouwenting@nbu.edu.cn}
\affiliation{Institute of Fundamental Physics and Quantum Technology, Department of Physics, School of Physical Science and Technology, Ningbo University, Ningbo, Zhejiang 315211, China}

\begin{abstract}
Applying the junction conditions to the slowly rotating Kerr spacetime with a thin shell, 
when higher order terms in the spin parameter $a$ can be neglected,
we find that while the angular momentum $L$ and Carter constant $C$ of the ray remain unchanged upon crossing the shell, its energy $E$ does not. 
Consequently, the impact parameters $\eta=L/E$ and $\xi=C/E^2$ of the ray are discontinued at the shell. 
Utilizing this transformation, we study the shadow of this spacetime and the corresponding images from an equatorial thin accretion disk.
The presence of the shell gives rise to distinctive features in the observed images. 
Notably, we observe distinct double photon rings in the images, which can gradually merge into a single ring.
Moreover, the shadow boundaries and the photon rings do not exhibit a one-to-one correspondence. 
The abrupt changes in redshift factor and the truncated photon regions profoundly influence the image, producing distinctive features such as the step-like structures.
These features in shell‑equipped spacetimes can help evaluate, through future astronomical observations, the applicability of the Israel junction condition and the shell model in real astrophysical systems.
\end{abstract}

\maketitle

\section{Introduction}\label{S1}
In general relativity, thin shell models are widely employed, where a thin shell serves as the interface connecting two distinct spacetime regions.
In the context of dark matter distributions, observations of dark matter halos forming high density spikes \cite{Eda:2013gg, Gondolo:1999ef, Gondolo:1999gy} provide indirect geometrical support for thin shell models.
The theoretical foundation for analyzing thin shells was established by Israel in 1967, who derived the relationship between the extrinsic curvature across a shell and its stress–energy tensor \cite{Israel:1966rt,MSM_1927__25__1_0}—now known as the Israel junction conditions. These conditions have since been applied in a variety of contexts, including notably in studies of cosmological phase transitions 
\cite{Blau:1986cw,Aurilia:1987cp,Berezin:1987bc}.
In such transitions, a new phase of matter grows as bubbles within an old, metastable phase. 
The boundary—or bubble wall—that separates these two phases with distinct properties and geometries can be modeled as a thin shell, to which the Israel junction conditions are directly applied.
Another well-known theoretical construct is the gravastar, which is characterized by an external spacetime resembling that of a black hole and an interior asymptotic de Sitter spacetime \cite{Mazur:2004fk,Chapline:2000en}.
Furthermore, the throat of the wormhole is also governed by the Israel junction condition. 
One can use it to study the properties of the throat as well as the stability of the wormhole \cite{Tsukamoto:2021fpp,Dutta:2025mxn}.

Since the Event Horizon Telescope (EHT) captured the first image of the black hole, the image of the black hole has become a powerful tool for us to probe the physics of strong gravitational fields \cite{EventHorizonTelescope:2019dse,EventHorizonTelescope:2022wkp}.
Recently, research on images of various compact objects, including black holes \cite{Mizuno:2018lxz,Uniyal:2025uvc,Peng:2020wun,Xu:2024gjs,Cao:2023par,Cao:2024kht,Cao:2024vtq,Sui:2023tje,Xiong:2025hjn,Meng:2023htc,Gan:2021xdl,Gan:2021pwu,Guo:2022ghl,Chen:2023qic,Zhang:2024lsf,Shen:2023nij,Zhang:2024jrw,Liang:2025bbn,Wang:2025dfn,Wang:2025edt,Bambi:2019tjh}, 
naked singularities \cite{Li:2025wmd,Shaikh:2019hbm,Gyulchev:2020cvo,Gyulchev:2021dvt,Chen:2023knf,Deliyski:2024wmt,Wu:2024ixf,Shaikh:2018lcc,HassanPuttasiddappa:2025tji,Huang:2024bbs,Vagnozzi:2022moj,Khodadi:2024ubi}, 
wormholes \cite{Huang:2023yqd,Chen:2024tss,Olmo:2021piq,Guerrero:2022qkh,Guo:2022iiy,Peng:2021osd,Luo:2023wru}, 
and other exotic compact objects \cite{Rosa:2022tfv,Huang:2025css,Tamm:2023wvn,Li:2025awg,Zhao:2025yhy,Guerrero:2022msp,Olmo:2023lil,Wang:2024uda} has attracted a lot of attention.
In \cite{Wang:2020emr,Cao:2025qzy,Koga:2025dii}, the authors utilize the junction conditions to study the behavior of rays passing through a wormhole throat. 
They find that while the photon's energy $E$ changes upon crossing the throat, its angular momentum $L$ remains conserved. 
This leads to a shift in the impact parameter $b$, resulting in a refractive-like effect for light traversing the throat.
Building on this framework, subsequent work \cite{Luo:2023wru, Peng:2021osd, Guo:2022iiy} further examines spherically symmetric wormhole images by explicitly incorporating this modified impact parameter.
In \cite{Cao:2025qzy}, we extend the analysis to Schwarzschild spacetime divided by a spherical shell—both static and dynamic—and demonstrate that the presence of such shells introduces several novel features into the resulting black hole images.

However, junction conditions are rarely considered in studies about images of rotating spacetimes.
In fact, the junction conditions for Kerr spacetime remain a challenging problem.
The counter-rotating Kerr spacetime separated by a shell is studied in \cite{Krisch:2009zs}.
Furthermore, in the case of a spacetime comprising a flat interior and a Kerr exterior joined by a thin shell, it has been shown that the angular velocity of the shell cannot be too large \cite{H_Pfister_1986}. 
Studies of thin shell models within slowly rotating Kerr backgrounds are also documented, including works such as \cite{DeLaCruz:1968zz,PhysRevD.18.1020}.
Based on the findings above, the authors examined the shadow of a slowly rotating Kerr black hole surrounded by a thin shell \cite{Qian:2021qow}. 
They observed novel features in the shadow, such as ``cuspy", which also appears in the shadows of rotating wormhole \cite{Gyulchev:2018fmd,Raza:2024frc}. 
For further studies on shadows of other rotating wormholes, see \cite{Gayen:2025tok}.
However, the shadow is formed solely by photons circling in the photon region.
As is well known, the black hole shadow is closely related to the structure of the photon ring, yet the photon ring itself only becomes observable when illuminated by an accretion disk.
This naturally raises the question: what would the image of a rotating spacetime with a thin shell look like when illuminated by a luminous accretion disk? 
Therefore, in this work, we investigate the optical appearance of a slowly rotating Kerr black hole enveloped by a shell.

In the slow-rotation approximation, 
we show that null rays preserve the angular momentum $ L $ and the Carter constant 
$C$ across the shell, 
whereas the energy $ E $ exhibits a discontinuous shift that induces a refraction-like transformation of the reduced impact parameters.
We obtain the critical curves from this transformation and compute the intensity profiles by backward ray tracing of a thin equatorial accretion disk.
We find that, similar to the thin-shell Schwarzschild spacetime, the shadow boundaries and photon rings in the thin-shell slowly rotating Kerr spacetime no longer exhibit a one-to-one correspondence. 
In addition, the presence of rotation substantially modifies the relations between photon rings in the image, 
and for certain specific configurations, two photon rings can merge into a single one, a phenomenon that does not occur in the thin shell Schwarzschild case. Furthermore, even though the shell itself is static, 
the redshift factor in the rotating thin shell spacetime remains discontinuous, which ultimately gives rise to step-like brightness features in the observed image.

This paper is organized as follows.
In Section \ref{S2}, we derive the transformations of the conserved quantities and the impact parameters for rays crossing the shell. 
Building on these results, Section \ref{S3} investigates the optical appearance of a slowly rotating Kerr spacetime surrounded by such a shell. 
Finally, we present our conclusions and discuss the implications of our findings in Section \ref{S4}.

\section{Thin shell junction of two Kerr spacetimes and transformation of conserved quantities}\label{S2}
In \cite{Cao:2025qzy}, we studied the junction of two Schwarzschild spacetimes connected by a thin shell.
We found that when photons traverse the shell, both the impact parameter and the redshift factor undergo discontinuous changes.
Besides, the four velocity of the photons is invariant, but their components change across the shell due to the shift in coordinates.
These transformations significantly affect the trajectories of light rays, giving rise to a variety of novel phenomenological effects.
Consequently, such observable features provide a potential means to test whether the Israel junction conditions and the thin shell model is physically viable in real astrophysical scenarios.
In this section, we will introduce the corresponding transformation of conserved quantities in the junction of two slowly rotating Kerr spacetimes.

We match two slowly rotating Kerr spacetimes by a rotating thin shell $\Sigma$ located at $r=R$.
The spacetimes outside and inside the thin shell (called exterior and interior spacetimes) are characterized by mass and spin parameters $(m_+, a_+)$ and $(m_-, a_-)$, respectively.
In the slow-rotation approximation, i.e., $(a/m)^2\ll1$, the Kerr metric reduces to the leading-order expression in $a$ , with all $\mathcal{O}\left(a^2\right)$ terms neglected \cite{DeLaCruz:1968zz,PhysRevD.18.1020,Qian:2021qow}:
\begin{eqnarray}\label{metricslow}
ds^2=-\left( 1-\frac{2m}{r} \right)dt^2+\frac{1}{ 1-\frac{2m}{r}}dr^2+r^2d\theta^2+r^2\sin^2\theta d\phi^2-\frac{4m a\sin^2\theta}{r}dt d\phi   \,.
\end{eqnarray}
It is shown that the shell is stable and can satisfy the relevant energy conditions \cite{Pereira:2023wnz}.
We ignore $\mathcal{O}\left(a^2\right)$ and higher-order terms in this and the following sections.
Let $(\tau, \theta, \psi)$ denote the intrinsic coordinates on the shell $\Sigma$.
The embeddings of $\Sigma$ into the exterior and interior spacetimes are specified by
\begin{eqnarray}
x_{\pm}^{\mu}(\tau,\theta,\psi)=\left(t_{\pm}(\tau),\,R,\,\theta,\,\phi_\pm(\psi,\,\tau)\right)
\end{eqnarray}
and they satisfy \cite{DeLaCruz:1968zz,PhysRevD.18.1020,Qian:2021qow}
\begin{eqnarray}
d\tau&=&\sqrt{f_\pm}dt_\pm    \,,\nonumber\\
d\psi&=&d\phi_\pm-\frac{2m_\pm a_\pm}{r^3}dt_\pm         \,,
\end{eqnarray}
where $f_\pm=1-2m_\pm/r$.
The angular coordinate basis vectors coincide on $\Sigma$ because the angular coordinates $\theta$ are identified.

We may also introduce a coordinate $\ell$ defined by $d\ell=1/\sqrt{f_\pm}dr$ with $\ell=0$ on $\Sigma$ and $\ell>0$ extending into the “$+$” side.
In terms of this coordinate and neglecting $a^2$, the metric can be expressed in Gaussian normal coordinates adapted to the shell as
\begin{eqnarray}\label{metricshell}
ds^2=-d\tau^2+d\ell^2+r^2d\theta^2+r^2\sin^2\theta d\psi^2   \,.
\end{eqnarray}
The transformation between the coordinate basis in \eqref{metricslow} and \eqref{metricshell} is given by
\begin{eqnarray}
\left( \begin{array}{c}
d\tau \\
d\ell \\
d\theta \\
d\psi
\end{array} \right)=
\left( \begin{array}{cccc}
\sqrt{f_\pm(R)} &0  &0  &0  \\
0   & \frac{1}{\sqrt{f_\pm(R)}}  &0   &0  \\
0  &0  & 1  &0   \\
-\frac{2m_\pm a_\pm}{R^3}    &0   &0   & 1
\end{array} \right)
\left( \begin{array}{c}
dt_\pm \\
dr_\pm \\
d\theta \\
d\phi_\pm
\end{array} \right)
\end{eqnarray}
and the dual basis are
\begin{eqnarray}
\left( \begin{array}{c}
\partial_\tau \\
\partial_\ell \\
\partial_\theta \\
\partial_\psi
\end{array} \right)
=\left( \begin{array}{cccc}
\frac{1}{\sqrt{f_\pm(R)}} &0  &0  & \frac{2m_\pm a_\pm}{R^3\sqrt{f_\pm(R)}}  \\
0   & \sqrt{f_\pm(R)}  &0   &0  \\
0  &0  & 1  &0   \\
0    &0   &0   & 1
\end{array} \right)
\left( \begin{array}{c}
\partial_{t_\pm} \\
\partial_{r_\pm} \\
\partial_\theta \\
\partial_{\phi_\pm}
\end{array} \right) \,.
\end{eqnarray}

This implies that the observers comoving with the shell coincide with zero angular momentum observers (ZAMO).
From these transformations, we can derive how the coordinate basis and their dual basis transform between the interior and exterior spacetimes:
\begin{eqnarray}
\left( \begin{array}{c}
dt_1 \\
dr_1 \\
d\theta \\
d\phi_1
\end{array} \right)
=\left( \begin{array}{cccc}
\sqrt{\frac{f_2(R)}{f_1(R)}} &0  &0  &0  \\
0   & \sqrt{\frac{f_1(R)}{f_2(R)}}  &0   &0  \\
0  &0  & 1  &0   \\
\sqrt{\frac{f_2(R)}{f_1(R)}}\frac{2m_1a_1}{R^3}-\frac{2m_2a_2}{R^3}    &0   &0   & 1
\end{array} \right)
\left( \begin{array}{c}
dt_2 \\
dr_2 \\
d\theta \\
d\phi_2
\end{array} \right)
\end{eqnarray}
and
\begin{eqnarray}
\left( \begin{array}{c}
\partial_{t_1} \\
\partial_{r_1} \\
\partial_{\theta} \\
\partial_{\phi_1}
\end{array} \right)
=\left( \begin{array}{cccc}
\sqrt{\frac{f_1(R)}{f_2(R)}} &0  &0  & \sqrt{\frac{f_1(R)}{f_2(R)}}\frac{2m_2a_2}{R^3}-\frac{2m_1a_1}{R^3}  \\
0   & \sqrt{\frac{f_2(R)}{f_1(R)}}  &0   &0  \\
0  &0  & 1  &0   \\
0    &0   &0   & 1
\end{array} \right)
\left( \begin{array}{c}
\partial_{t_2} \\
\partial_{r_2} \\
\partial_{\theta} \\
\partial_{\phi_2}
\end{array} \right)  \,,
\end{eqnarray}
where $(1,2)$ represents $(+, -)$ or $(-, +)$.

The Killing vectors of the exterior and interior spacetimes are $K_{t_\pm}=\partial_{t_\pm}$ and $K_{\phi_\pm}=\partial_{\phi_\pm}$.
Since the four-vector of the photon $k^a$ remains continuous across the shell \cite{Cao:2025qzy}, the angular momentum $L$ of the photons does not change when they passage through the shell,
\begin{eqnarray}
L_1=(\partial_{\phi_1})^ak_a=(\partial_{\phi_2})^ak_a=L_2\equiv L   \,.
\end{eqnarray}
However, the energy $E$ transforms as
\begin{eqnarray}\label{E}
E_1&=&-(\partial_{t_1})^ak_a   \nonumber\\
&=&-\sqrt{\frac{f_1(R)}{f_2(R)}}(\partial_{t_2})^ak_a-\left( \sqrt{\frac{f_1(R)}{f_2(R)}}\frac{2m_2a_2}{R^3}-\frac{2m_1a_1}{R^3} \right)(\partial_{\phi_2})^ak_a    \nonumber\\
&=&\sqrt{\frac{f_1(R)}{f_2(R)}}E_2-\left( \sqrt{\frac{f_1(R)}{f_2(R)}}\frac{2m_2a_2}{R^3}-\frac{2m_1a_1}{R^3} \right)L       \,.
\end{eqnarray}
The null geodesics are entirely determined by the impact parameter $\eta=L/E$ and the reduced Carter constant $\xi=C/E^2$, where $C$ is the Carter constant.
Hence $\eta$ transforms as
\begin{eqnarray}\label{eta}
\frac{1}{\eta_1}=\sqrt{\frac{f_1(R)}{f_2(R)}}\frac{1}{\eta_2}-\left( \sqrt{\frac{f_1(R)}{f_2(R)}}\frac{2m_2a_2}{R^3}-\frac{2m_1a_1}{R^3} \right)       \,.
\end{eqnarray}
The photon’s four velocity is invariant across the shell.
The null geodesics in the Kerr spacetime are described by
\begin{eqnarray}
k_\theta^2=C+\cos^2\theta\left( a^2E^2-\frac{L^2}{\sin^2\theta} \right)  \,.
\end{eqnarray}
Therefore, $C$ is invariant across the shell when $\mathcal{O}(a^2)$ terms are neglected:
\begin{eqnarray}
C_+&\approx& k_{\theta+}^2+\frac{\cos^2\theta}{\sin^2\theta}L^2  \nonumber\\
&\approx&\left( k_a(\partial_\theta)^a \right)^2+\frac{\cos^2\theta}{\sin^2\theta}L^2   \nonumber\\
&\approx& k_{\theta-}^2+\frac{\cos^2\theta}{\sin^2\theta}L^2   \nonumber\\
&\approx& C_-  \,.
\end{eqnarray}
Although here $C$ is invariant only after neglecting $\mathcal{O}(a^2)$, it is worth emphasizing that our method does not rely on the slow-rotation approximation.
Once the junction conditions  are determined, i.e., the coordinate transformation between the interior and exterior spacetimes, our approach can be used to compute the change in $C$, regardless of whether the $a^2$ terms are neglected or not.
Of course, in that case $C$ may turn out to vary, and likely depends on the energy $E$ and spin $a$.

As a result, the transformation of $\xi$ is given by
\begin{eqnarray}\label{xi}
\xi_1=\frac{\xi_2}{\left[\sqrt{\frac{f_1(R)}{f_2(R)}}-\left(\sqrt{\frac{f_1(R)}{f_2(R)}}\frac{2m_2a_2}{R^3}-\frac{2m_1a_1}{R^3}  \right)\eta_2\right]^2}   \,.
\end{eqnarray}

In this section, we have studied the transformation of Killing conserved quantities as photons cross the thin shell.
We find that while the angular momentum $L$ and the Carter constant $C$ remain invariant across the shell, the energy $E$ transforms according to \eqref{E}.
This discontinuity in $E$ , in turn, induces transformations in $\eta$ and $\xi$ , given by \eqref{eta} and \eqref{xi}, respectively.
These changes lead to a refraction of rays and result in a jump in the redshift factor at the shell.
Collectively, these effects significantly influence photon trajectories and leave a distinct imprint on the observed black hole image.

\section{Shadow and image of Kerr spacetime with a thin shell}\label{S3}
The null geodesics of the slowly rotating Kerr spacetime can be obtained by neglecting $\mathcal{O}\left(a^2\right)$ terms in the full Kerr null geodesic equations:
\begin{eqnarray}
\dot{t}&=& \frac{1}{f}-\frac{2ma\eta}{r^3f}     \,,\nonumber\\
\dot{\phi}&=&\frac{\eta}{r^2\sin^2\theta}-\frac{a}{r^2}+\frac{a}{r(r-2m)}   \,,\nonumber\\
r^4\dot{r}^2&=& r^4-r^2f(\xi+\eta^2)-4am\eta r\equiv \mathcal{R}(r)   \,,\nonumber\\
r^4\dot{\theta}^2&=& \xi-\eta^2\cot^2\theta  \equiv \Theta(\theta)      \,.
\end{eqnarray}
For photons orbiting in the photon region $(r_{min}, r_{max})$, $\eta$ and $\xi$ are determined by solving the conditions $\mathcal{R}(r) =\mathcal{R}'(r) =0$:
\begin{eqnarray}
\eta&=&\frac{-r^2(r-3m)}{2am}   \,,\nonumber\\
\xi&=&3r^2-\eta^2        \,.
\end{eqnarray}
Solving $\xi=0$  yields two solutions that define the boundaries of the photon region, $(r_{min}, r_{max})$, as
\begin{eqnarray}
r_{max,min}=3m\pm \frac{2\sqrt{3}}{3}a +\mathcal{O}\left(a^2\right)  \,.
\end{eqnarray}
In the limit $a\rightarrow0$ and $r\in(3m- \frac{2\sqrt{3}}{3}a,3m+\frac{2\sqrt{3}}{3}a)$, this result agrees with the corresponding Kerr case.

A photon emitted from the photon region with impact parameters $(\eta, \xi)$ and observed by a distant observer will be mapped to the point $(x,y)$ in the celestial coordinates, where
\begin{eqnarray}
x&=&-\frac{\eta}{\sin\theta_o}     \,,\nonumber\\
y&=&\pm\sqrt{\xi-\eta^2\cot^2\theta_o}   \,,
\end{eqnarray}
and $\theta_o$ is defined as the inclination angle of the observer.
We set $\theta_o=\pi/2$
in the following.
It should be noted that when photons cross the shell, their impact parameters $(\eta, \xi)$ undergo a transformation.
Therefore, for photons originating from the photon region inside the shell, $(\eta, \xi)$ must be transformed before being substituted into this formula.

We adopt a thin accretion disk model confined to the equatorial plane.
This model has been widely used in previous studies of black hole images \cite{Wang:2025edt,Wang:2024uda}. 
Its main feature is that the full hydrodynamical evolution of the accreting matter is not solved. 
Instead, the motion of the emitting matter is prescribed by the timelike geodesics of the background spacetime. 
In this disk model, whose explicit velocity prescription will be specified below, the four-velocity of the emitting matter is strongly controlled by the background geometry.
For the thin-shell spacetime considered in this work, the two sides of the shell are described by different Kerr geometries. 
Thus, when the same disk model is applied on both sides of the shell, the four-velocity of the emitting matter near the shell generally changes. 
In this sense, the discontinuity of the disk four-velocity is not introduced as an arbitrary additional structure. 
Rather, it follows from applying the analytic disk model to a thin-shell spacetime, where the velocity prescription depends on the background geometry.

At the same time, we emphasize that the present work does not provide a complete dynamical model for massive matter crossing the shell. 
Unlike the propagation of photons, massive accreting matter may interact with the shell through collisions, energy dissipation, or momentum transfer when crossing it. 
Therefore, the four-velocity of matter on the two sides of the shell is not necessarily continuous. 
The shell may also be interpreted as a physical interface or phase boundary, such as one arising in a cosmological phase transition. 
In this case, the matter in the interior and exterior regions may originate from different accretion histories and hence possess different dynamical properties. 
Therefore, within the simplified disk model adopted in this work, we allow the four-velocity of the emitting matter to differ across the shell.

The velocity of the accretion disk matter on either side of the shell is as follows.
Outside the innermost stable circular orbit (ISCO), matter follows stable circular orbits.
Inside the ISCO, the material free infalls along geodesic trajectories, with the energy, angular momentum and Carter constant set to the values of material at the ISCO radius. 
When the inner ISCO lies outside the shell, the parameters of material slightly outside the shell are forbidden for free-falling orbits in the interior spacetime. 
Consequently, we adopt the parameters of the inner ISCO for the matter in the interior spacetime, even though this ISCO lies outside the shell and hence does not physically exist.
Therefore, the four velocity of the accreting matter on the two sides near the shell are different in our prescription.
The redshift factor is defined by
\begin{eqnarray}
g_i=\frac{\omega_{obs}}{\omega_{em}}=\frac{u_{obs}^ak_a}{u_{em}^ak_a}   \,,
\end{eqnarray}
where $\omega_{obs}$ and $\omega_{em}$ are the frequency measured by the distant observer and the luminous material.
$u_{obs}$ is the four velocity of the ZAMO observer, and $u_{em}$ is the four velocity of accreting matter, which is either circular geodesic motion or free-fall.
Since the redshift factor depends on the velocity of the luminous matter, this discontinuity leads directly to a jump in the redshift at the shell.
This, in turn, gives rise to step-like discontinuities in the observed intensity.

In this paper, we refer to the shadow boundary (also known as the critical curve) as the image formed by photons associated with orbits in the photon region. In typical black hole images, this curve often corresponds to the photon ring in the image produced by an accretion disk. In the presence of a thin shell, however, this correspondence can be broken.
In the following, we present several key shadows and images illustrating the influence of the junction under different configurations.
The emitted intensity is \cite{Hou:2022eev}
\begin{eqnarray}
I_{em}(r)=e^{-\frac{1}{2}\left( \ln\frac{r}{r_{h}} \right)^2-2\ln\frac{r}{r_{h}}}   \,,
\end{eqnarray}
where $r_h$ is outer horizon of the interior spacetime.

In the following, we use several representative examples to illustrate some characteristic features introduced by the thin shell.
First, we examine the relation between the two photon rings in the image, which can merge for specific configurations.
Second, we discuss the step-like brightness features caused by the discontinuity of the redshift factor across $\Sigma$.
Finally, we show that ring-like structures can persist even when the photon region is truncated,
thereby breaking the one-to-one correspondence between shadow boundaries and photon rings.

\subsection{$m_-=0.6,a_-=0.3,m_+=1.13,a_+=0.01,R=2.7$.}

\begin{figure}[htbp]
		\begin{minipage}[t]{0.45\linewidth}
			\centering
			\includegraphics[height=6cm]{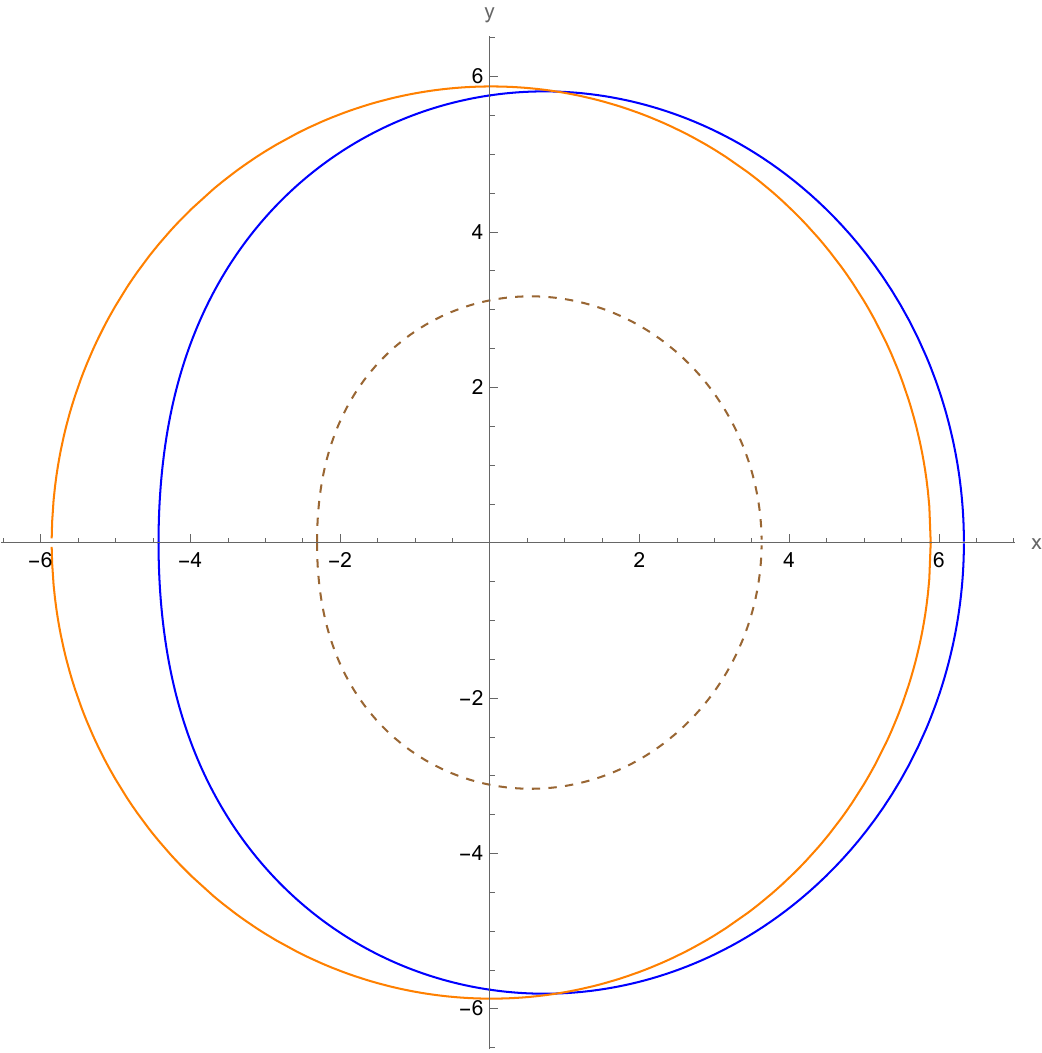}
        \par\vspace{1mm}
        \small{(a)}
		\end{minipage}%
		\begin{minipage}[t]{0.45\linewidth}
			\centering
			\includegraphics[height=6cm]{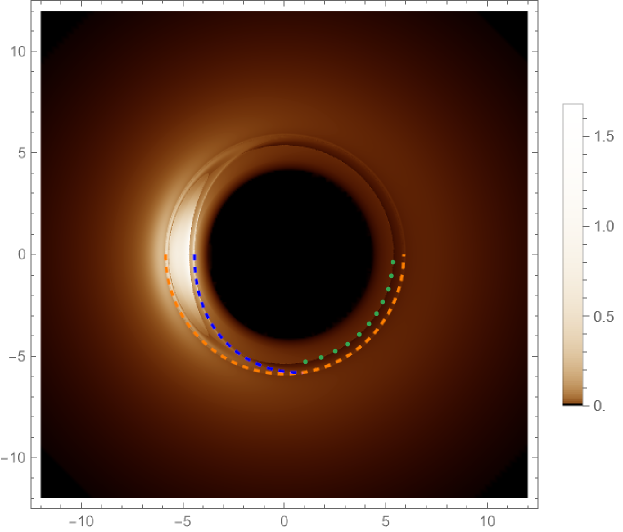}
        \par\vspace{1mm}
        \small{(b)}
		\end{minipage}%

		\begin{minipage}[t]{0.45\linewidth}
			\centering
			\includegraphics[height=6cm]{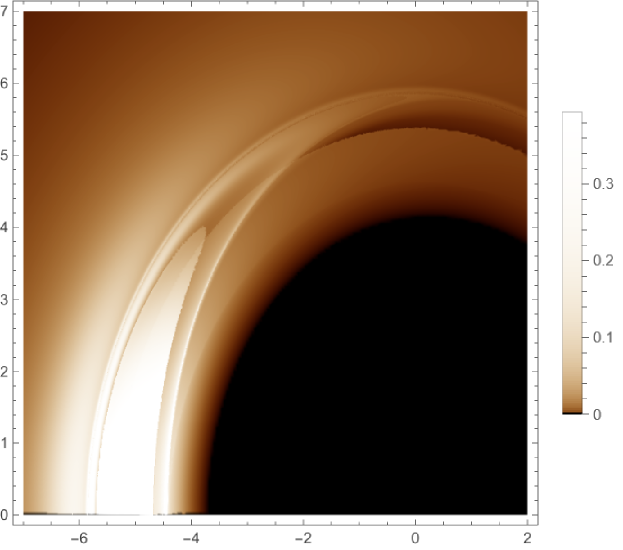}
        \par\vspace{1mm}
        \small{(c)}
		\end{minipage}%
		\begin{minipage}[t]{0.45\linewidth}
			\centering
			\includegraphics[height=6cm]{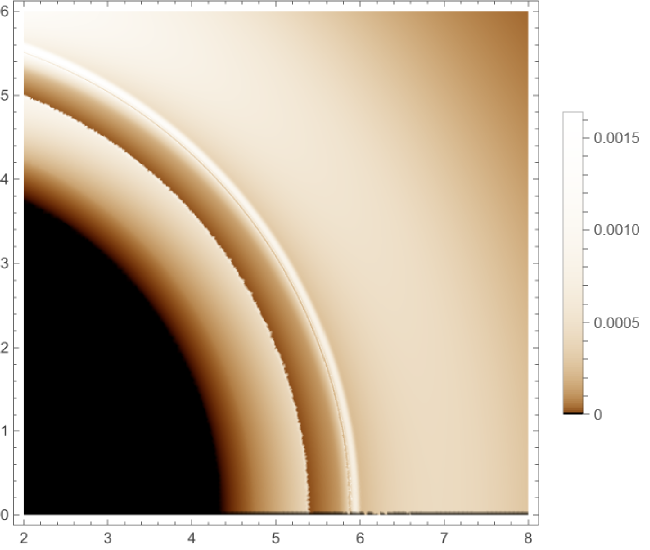}
        \par\vspace{1mm}
        \small{(d)}
		\end{minipage}%

  \caption{The shadow (a), images (b) and enlarged view of the image (c) and (d) of slowly rotating Kerr black hole with $m_-=0.6,a_-=0.3,m_+=1.13,a_+=0.01,R=2.7$.
  In (a), the orange curve represents the shadow boundary corresponds to the outer photon region and the blue one corresponds to the inner photon region. 
  The brown dashed curve represents the shadow boundary corresponds to the inner photon region when no thin shell is present.
  In (b), the dashed orange and blue curve represents the outer and inner photon ring, respectively.
  The green dashed curves indicates the ``step" feature resulting from a discontinuity in the redshift factor.
  }
  \label{m0.6}
\end{figure}

In this case, the inner and outer photon region are $r_{PR-}\subseteq(1.33,2.10)$ and $r_{PR+}\subseteq(3.38,3.40)$, respectively.
Since both photon regions are complete, the shadow boundaries (critical curve) are represented by solid curves.
The brown dashed curve represents the shadow boundary corresponds to the inner photon region in the absence of a shell. 
In contrast, its counterpart in the shell-present case (blue curve) is visibly larger, demonstrating the shell's enlarging  effect.
As shown in  Fig. \ref{m0.6} (b), there are still two photon rings on the left side.
The orange curve depicts the outer photon ring and the blue one depicts the inner photon ring.
However, they gradually merge into a single structure, and on the right side, only the outer photon ring remains.
The reason is as follows: In Fig. \ref{m0.6} (a), when the blue circle lies outside the orange circle, it means that the peak of the effective potential of the inner photon region is lower than that of the outer photon region. 
As a result, the photon ring associated with the inner photon region is nonexistent, and only the photon ring from the outer photon region can be observed.
On the other hand, in addition to the photon rings, there is a ``step" (depicted by the green curve) on the right side in  Fig. \ref{m0.6} (b), which results from the discontinuity in the redshift factor at the shell, as we discussed earlier.
This step-like feature is a direct consequence of the velocity discontinuity of the accretion disk matter at the shell.

\begin{figure}[htbp]
		\begin{minipage}[t]{\linewidth}
			\centering
			\includegraphics[height=5cm]{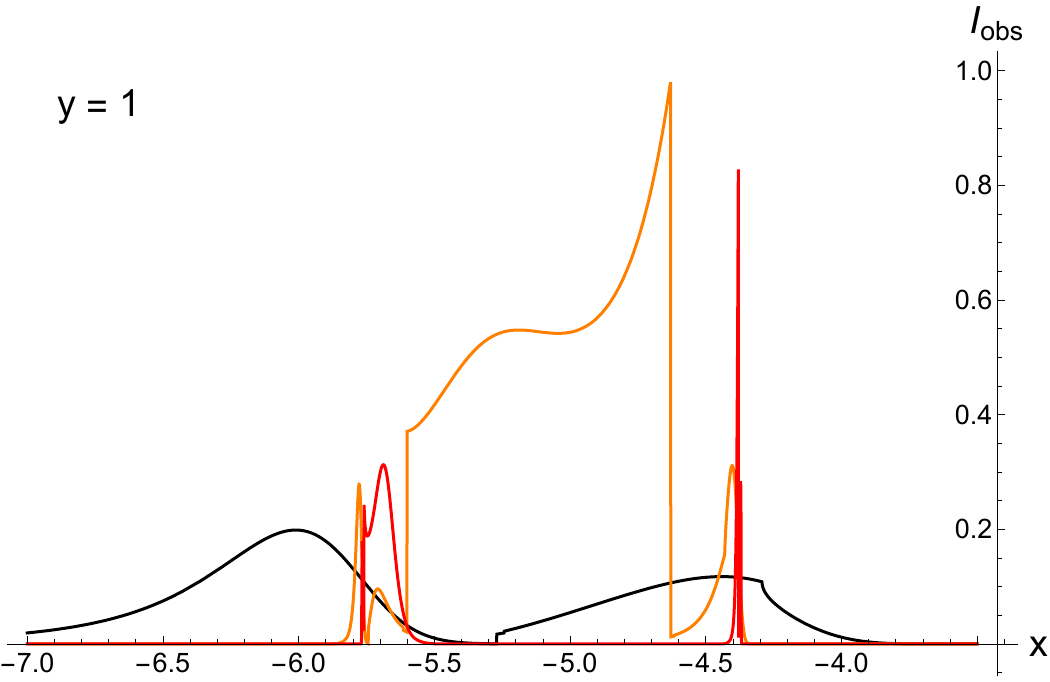}
        \par\vspace{1mm}
        \small{(a)}
		\end{minipage}%

		\begin{minipage}[t]{\linewidth}
			\centering
			\includegraphics[height=8cm]{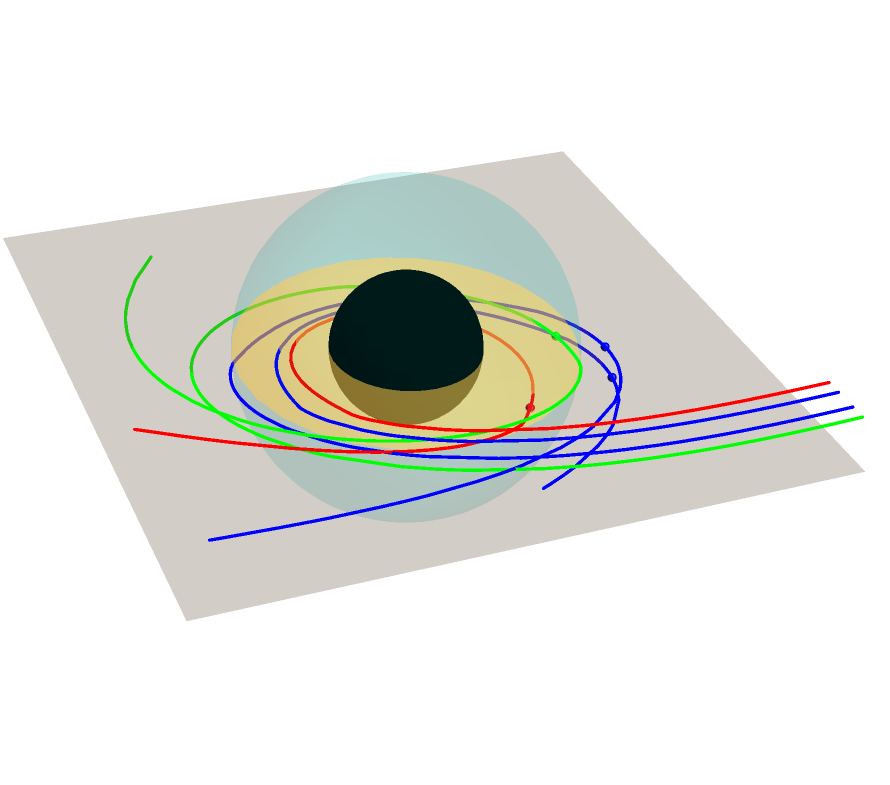}
        \par\vspace{1mm}
        \small{(b)}
		\end{minipage}%

  \caption{(a) The observed intensity of $y=1$.
  The black, orange and red curves correspond to the intensity contributions from rays intersecting the accretion disk for the first,  second and third time, respectively.
  (b) The trajectory of rays that intersect the accretion disk for the second time.
  The light blue sphere represents the shell. The orange plane denotes the accretion disk inside the shell, while the gray plane represents the accretion disk outside the shell.
  Blue rays represent intersections occurring outside the spherical shell, and red and green rays represent intersections inside the shell.
 }
  \label{m0.6tra}
\end{figure}

We can also observe a bright spot between the two photon rings on the left side of Fig. \ref{m0.6} (b), which is produced by rays that intersect the accretion disk for the second time (lensed). 
The detailed mechanism is illustrated in Fig. \ref{m0.6tra}.
As shown in  Fig. \ref{m0.6tra} (a), for $x\subseteq(-5.60,-4.63)$, 
the intensity of the orange component increases significantly. 
This is because the rays in this region, whose trajectories intersect the accretion disk for the second time at points outside the spherical shell—such as the blue rays in Fig. \ref{m0.6tra} (b)—experience less redshift. 
In contrast, rays with $x\lesssim -5.60$ or $x\gtrsim -4.63$ intersect the disk for the second time inside the shell, as illustrated by the green and red rays in  Fig. \ref{m0.6tra} (b). 
Since the redshift factor exhibits a discontinuity across the shell, these rays undergo a sudden change in redshift, leading to a corresponding discontinuity in their observed intensity. 
This results in a sharp brightness enhancement in the interval $(-5.60,-4.63)$.

\begin{figure}[htb]
    \centering
    \includegraphics[width=6cm]{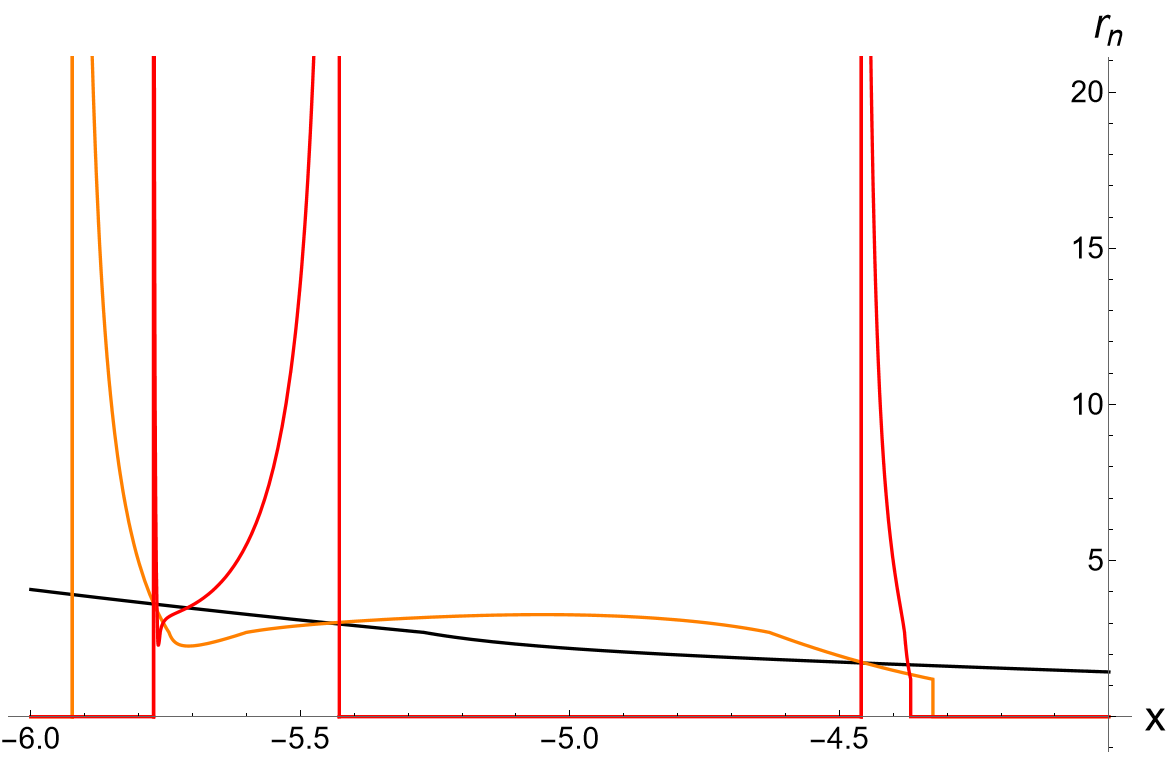}
    \caption{The transfer function at $y=1$.
    The black, orange and red curves represent the transfer functions for rays intersecting the accretion disk for the first, second, and third times, respectively.  }
    \label{transfer}
\end{figure}

Finally, a V-shaped pattern in the transfer functions is also observed here, analogous to that of a Schwarzschild spacetime with a shell, as shown in Fig. \ref{transfer}.

\subsection{$m_-=1,a_-=0.3,m_+=1.15,a_+=0.2,R=3.25$.}

\begin{figure}[htbp]
		\begin{minipage}[t]{0.45\linewidth}
			\centering
			\includegraphics[height=6cm]{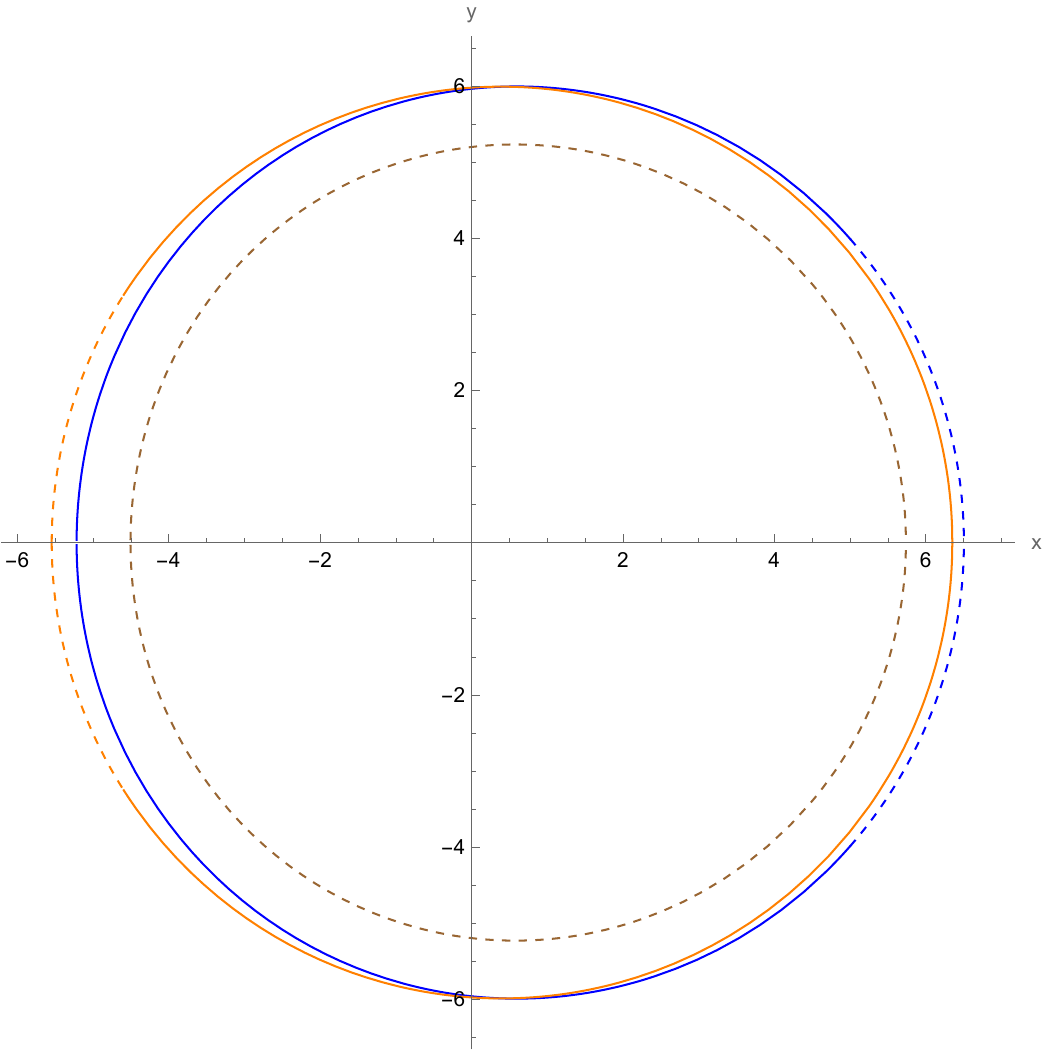}
        \par\vspace{1mm}
        \small{(a)}
		\end{minipage}%
		\begin{minipage}[t]{0.45\linewidth}
			\centering
			\includegraphics[height=6cm]{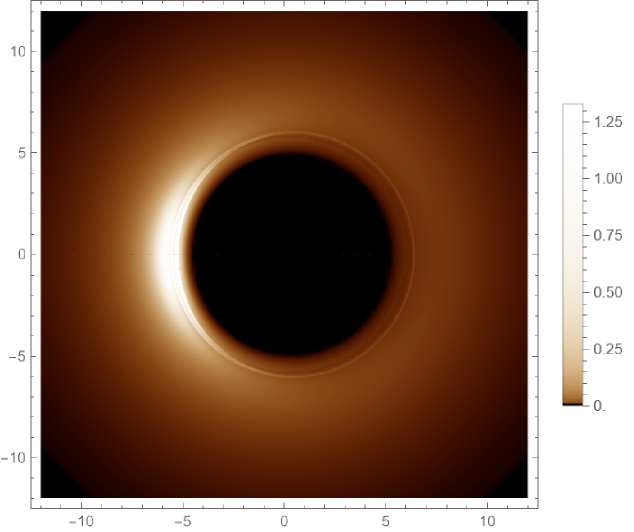}
        \par\vspace{1mm}
        \small{(b)}
		\end{minipage}%

		\begin{minipage}[t]{0.45\linewidth}
			\centering
			\includegraphics[height=6cm]{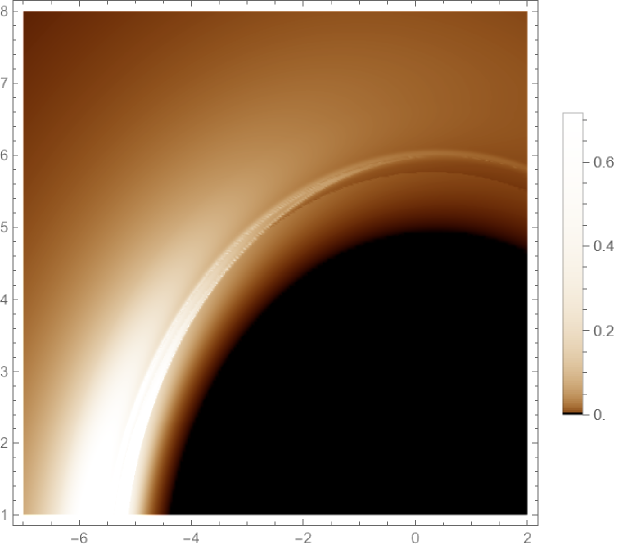}
        \par\vspace{1mm}
        \small{(c)}
		\end{minipage}%
		\begin{minipage}[t]{0.45\linewidth}
			\centering
			\includegraphics[height=6cm]{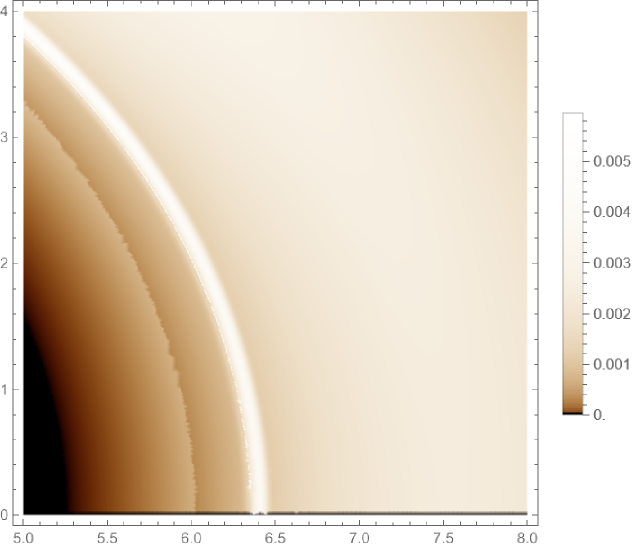}
        \par\vspace{1mm}
        \small{(d)}
		\end{minipage}%

  \caption{The shadow (a), images (b) and enlarged view of the image (c) and (d) of slowly rotating Kerr black hole with $m_-=1,a_-=0.3,m_+=1.15,a_+=0.2,R=3.25$.
  In (a), the orange curve represents the shadow boundary corresponds to the outer photon region and the blue one corresponds to the inner photon region.
  The dashed curve represents the hypothetical shadow that would exist if it were not truncated by the shell — a shadow that does not actually form due to the interruption caused by the shell.
   The brown dashed curve represents the shadow boundary corresponds to the inner photon region when no thin shell is present.
}
  \label{m1a0.4}
\end{figure}

The photon regions are identified at $r_{PR-}\subseteq(2.60,3.25)$ for the interior and $r_{PR+}\subseteq(3.25,3.67)$ for the exterior, respectively.
Fig. \ref{m1a0.4} (b) is similar to Fig. \ref{m0.6} (b).
On the left side of both plots, two photon rings are present, gradually merging into a single ring toward the right.
On the right side of Fig. \ref{m1a0.4} (a), the blue circle lies outside the orange one, which explains why only a single outer photon ring appears on the right side of Fig. \ref{m1a0.4} (b).
Similarly, bright spots and ``steps"  are also present in the image.
However, unlike Fig. \ref{m0.6} (a), the orange circle on the left side of Fig. \ref{m1a0.4} (a) is dashed, indicating that the corresponding shadow is truncated by a shell.
In contrast, in Fig. \ref{m1a0.4} (b), the photon ring associated with the left side region remains visible.
This occurs because, according to ray-tracing method, rays are traced backward from the observer toward the source.
When these rays approach the location of the truncated outer photon region (or the shell), they have already orbited the black hole many times before reaching that radius.
During these multiple orbits, the rays intersect the accretion disk repeatedly, and each intersection contributes to the total observed intensity.
Even though the outer photon region is truncated by the spherical shell—meaning that rays cannot complete their infinite winding motion asymptotically approaching a constant radius and instead transition into the interior spacetime upon reaching the shell—the reduction in total observed intensity is relatively minor.
This is because the observable flux comes not only from the final, infinitesimal winding near the exact photon orbit, but also from the repeated crossings of the accretion disk by rays close to the critical impact parameter.
As a result, even with the truncation, sufficient light is collected along the trajectory to produce a visible ring-like feature in the image.
Thus, a photon ring or a lensed ring still appears at the position corresponding to the (now truncated) outer photon region.
This demonstrates that a ring-like brightness enhancement can survive even when the corresponding photon region is truncated, hence photon rings and critical curves are not in one-to-one correspondence in thin shell spacetimes.
A more detailed explanation can be found in our previous work \cite{Cao:2025qzy}.

\subsection{$m_-=0.5,a_-=0.01,m_+=1.4,a_+=0.2,R=4.3$.}

\begin{figure}[htbp]
		\begin{minipage}[t]{0.45\linewidth}
			\centering
			\includegraphics[height=6cm]{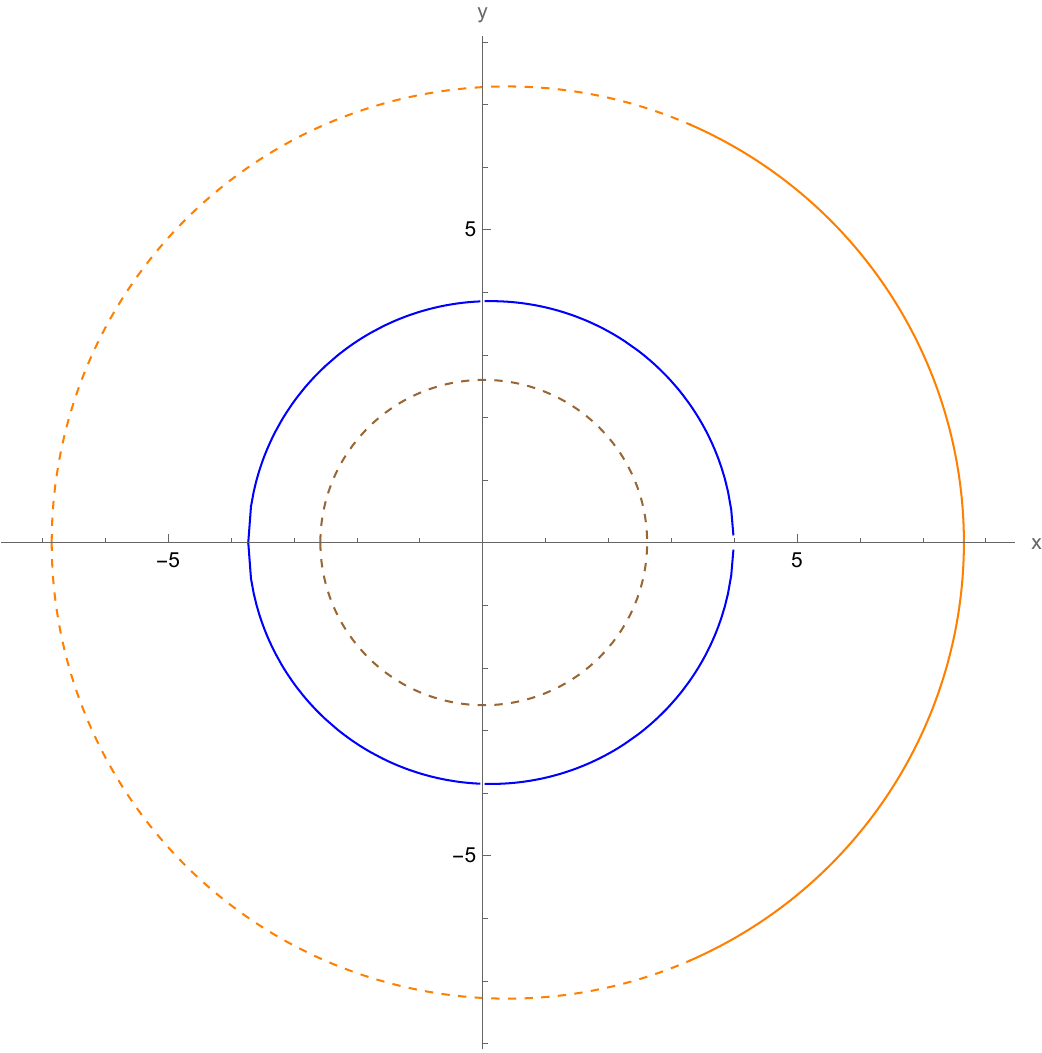}
        \par\vspace{1mm}
        \small{(a)}
		\end{minipage}%
		\begin{minipage}[t]{0.45\linewidth}
			\centering
			\includegraphics[height=6cm]{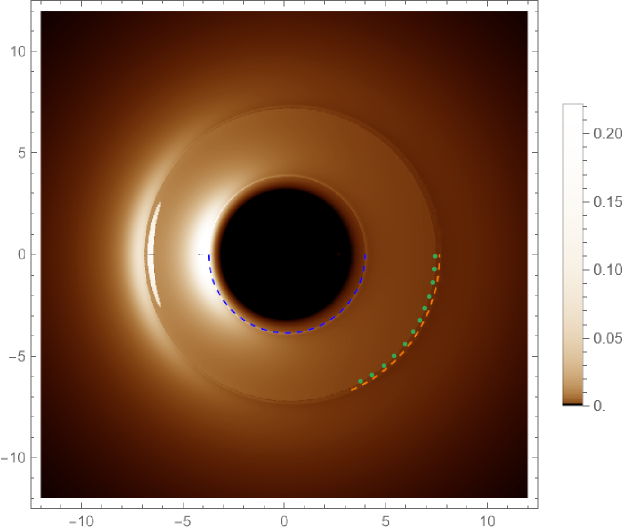}
        \par\vspace{1mm}
        \small{(b)}
		\end{minipage}%

		\begin{minipage}[t]{0.45\linewidth}
			\centering
			\includegraphics[height=6cm]{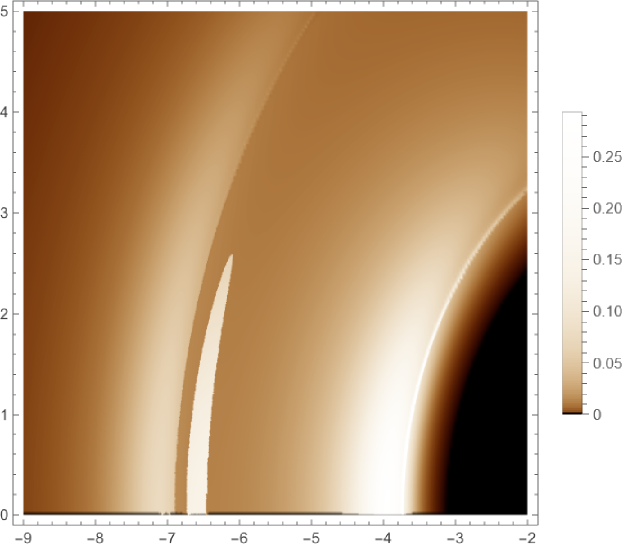}
        \par\vspace{1mm}
        \small{(c)}
		\end{minipage}%
		\begin{minipage}[t]{0.45\linewidth}
			\centering
			\includegraphics[height=6cm]{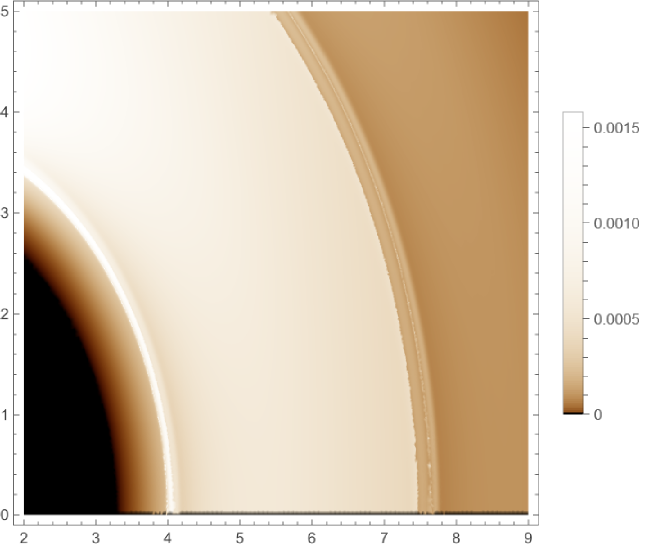}
        \par\vspace{1mm}
        \small{(d)}
		\end{minipage}%

  \caption{The shadow (a), images (b) and enlarged view of the image (c) and (d) of slowly rotating Kerr black hole with $m_-=0.5,a_-=0.01,m_+=1.4,a_+=0.2,R=4.3$.
  In (a), the orange curve represents the shadow boundary corresponds to the outer photon region and the blue one corresponds to the inner photon region. 
  The orange dashed curve represents the hypothetical shadow that would exist if it were not truncated by the shell.
   The brown dashed curve represents the shadow boundary corresponds to the inner photon region when no thin shell is present.
     In (b), the dashed orange and blue curve represents the outer and inner photon ring, respectively.
     The green dashed curves indicates the ``step" feature resulting from a discontinuity in the redshift factor.}
  \label{m0.5}
\end{figure}

The photon regions are identified at $r_{PR-}\subseteq(1.49,1.51)$ for the interior and $r_{PR+}\subseteq(4.30,4.42)$ for the exterior, respectively. 
While the inner photon region is complete, the outer photon region is truncated by the shell. 
Consequently, only the solid orange curve in  Fig. \ref{m0.5} (a) corresponds to the shadow cast by the outer photon ring.
In Fig. \ref{m0.5} (b), there is a complete inner photon ring (dashed blue curve), while the outer photon ring (dashed orange curve) appears only on the right side and is thus incomplete. 
This is consistent with the shadow in Fig. \ref{m0.5} (a).
As noted previously, the ``step" (dashed green curve) due to the discontinuous redshift factor is still present. 
In addition, a bright spot is observed.

\subsection{$m_-=1,a_-=0.3,m_+=1.33,a_+=0.01,R=2.7$.}

\begin{figure}[htbp]
		\begin{minipage}[t]{0.45\linewidth}
			\centering
			\includegraphics[height=6cm]{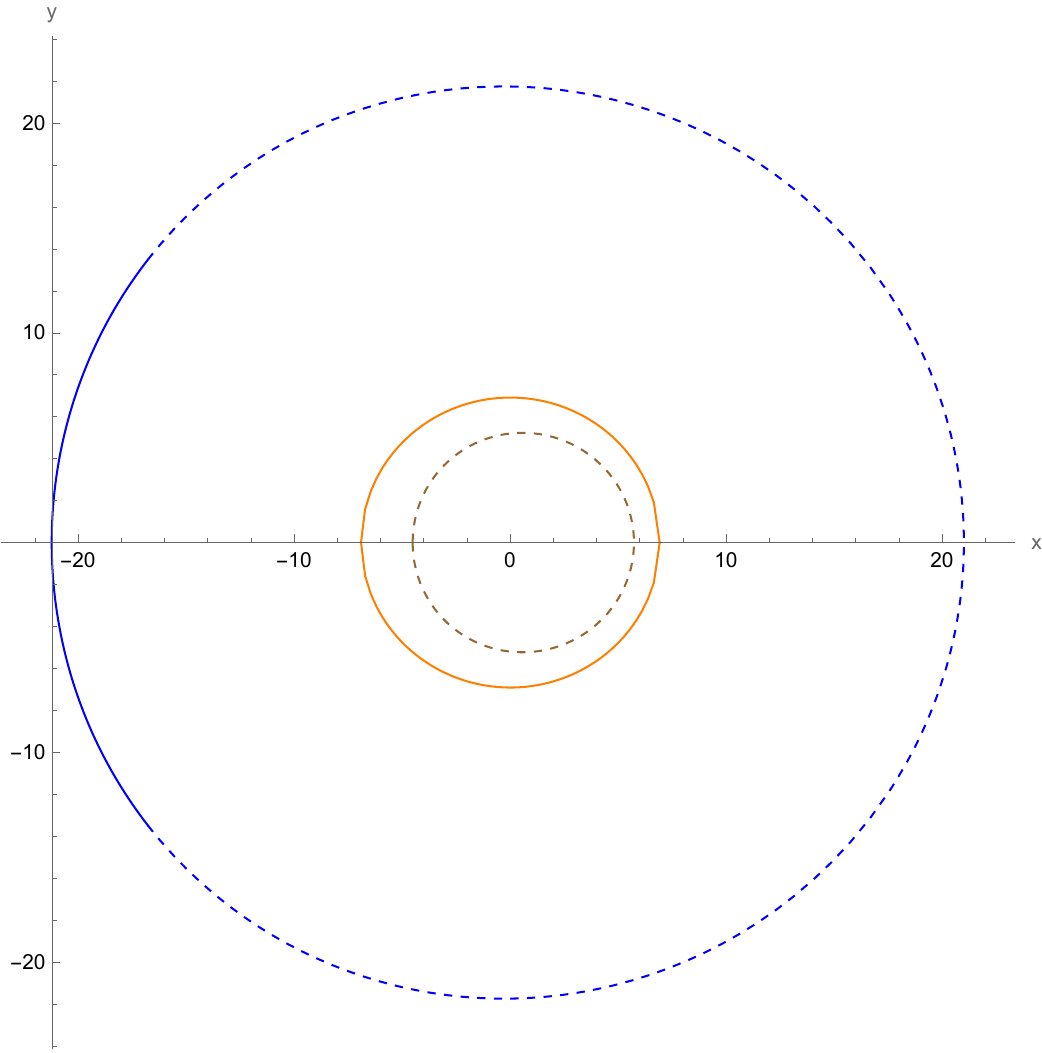}
        \par\vspace{1mm}
        \small{(a)}
		\end{minipage}%
		\begin{minipage}[t]{0.45\linewidth}
			\centering
			\includegraphics[height=6cm]{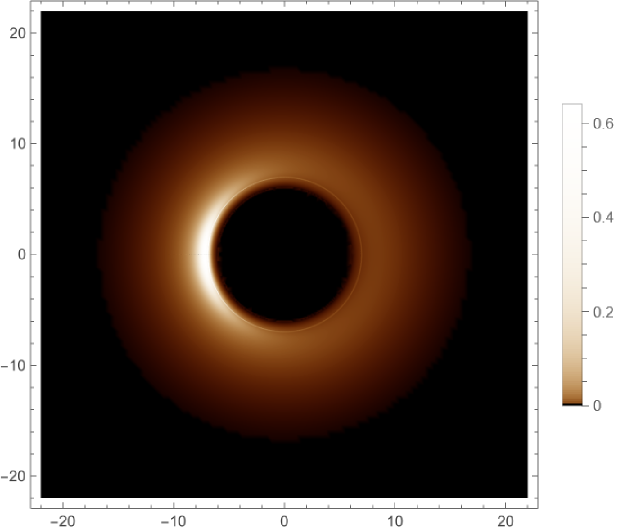}
        \par\vspace{1mm}
        \small{(b)}
		\end{minipage}%

		\begin{minipage}[t]{0.45\linewidth}
			\centering
			\includegraphics[height=6cm]{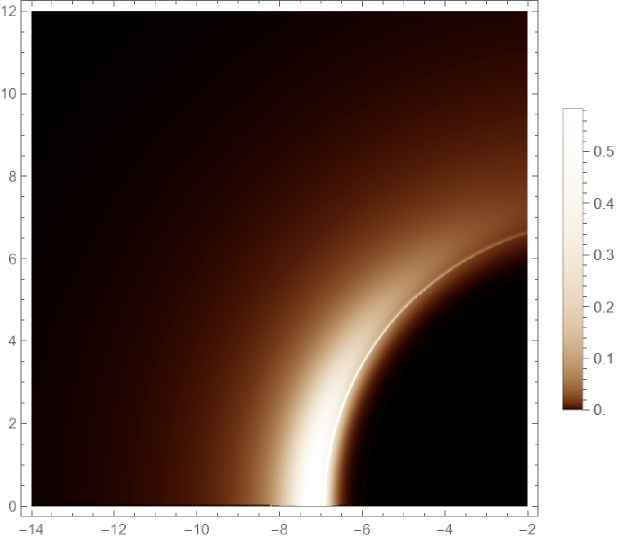}
        \par\vspace{1mm}
        \small{(c)}
		\end{minipage}%
		\begin{minipage}[t]{0.45\linewidth}
			\centering
			\includegraphics[height=6cm]{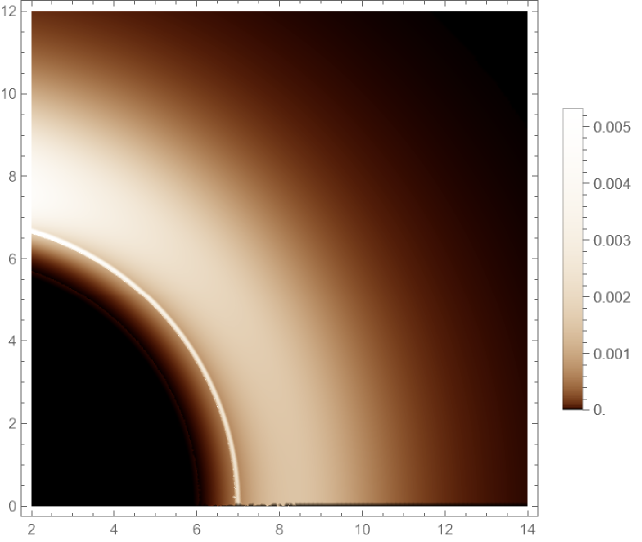}
        \par\vspace{1mm}
        \small{(d)}
		\end{minipage}%

  \caption{The shadow (a), images (b) and enlarged view of the image (c) and (d) of slowly rotating Kerr black hole with $m_-=1,a_-=0.3,m_+=1.33,a_+=0.01,R=2.7$.
  The orange curve represents the shadow boundary corresponds to the outer photon region and the blue one corresponds to the inner photon region.
  The blue dashed curve represents the hypothetical shadow that would exist if it were not truncated by the shell.
   The brown dashed curve represents the shadow boundary corresponds to the inner photon region when no thin shell is present.}
  \label{m1}
\end{figure}

The photon regions are identified at $r_{PR-}\subseteq(2.60,2.70)$ for the interior and $r_{PR+}\subseteq(3.98,4.00)$ for the exterior, respectively. 
As illustrated in Fig. \ref{m1} (a), the blue circle lies outside the orange circle, indicating that the peak value of the effective potential in the outer photon region is higher than that in the inner photon region. 
Consequently, no photon ring corresponding to the inner photon region is formed, resulting in only a single outer photon ring being visible in Fig. \ref{m1} (b). 
As previously mentioned, there are still bright spots near the photon ring and the associated ``step". 
However, in this case, the redshift factor at the ``step" is extremely small—approximately $10^{-2}$—leading to a final intensity too faint to be discernible in the image.

\subsection{$m_-=0.5,a_-=0.01,m_+=1.4,a_+=0.2,R=3.8$.}

\begin{figure}[htbp]
		\begin{minipage}[t]{0.45\linewidth}
			\centering
			\includegraphics[height=6cm]{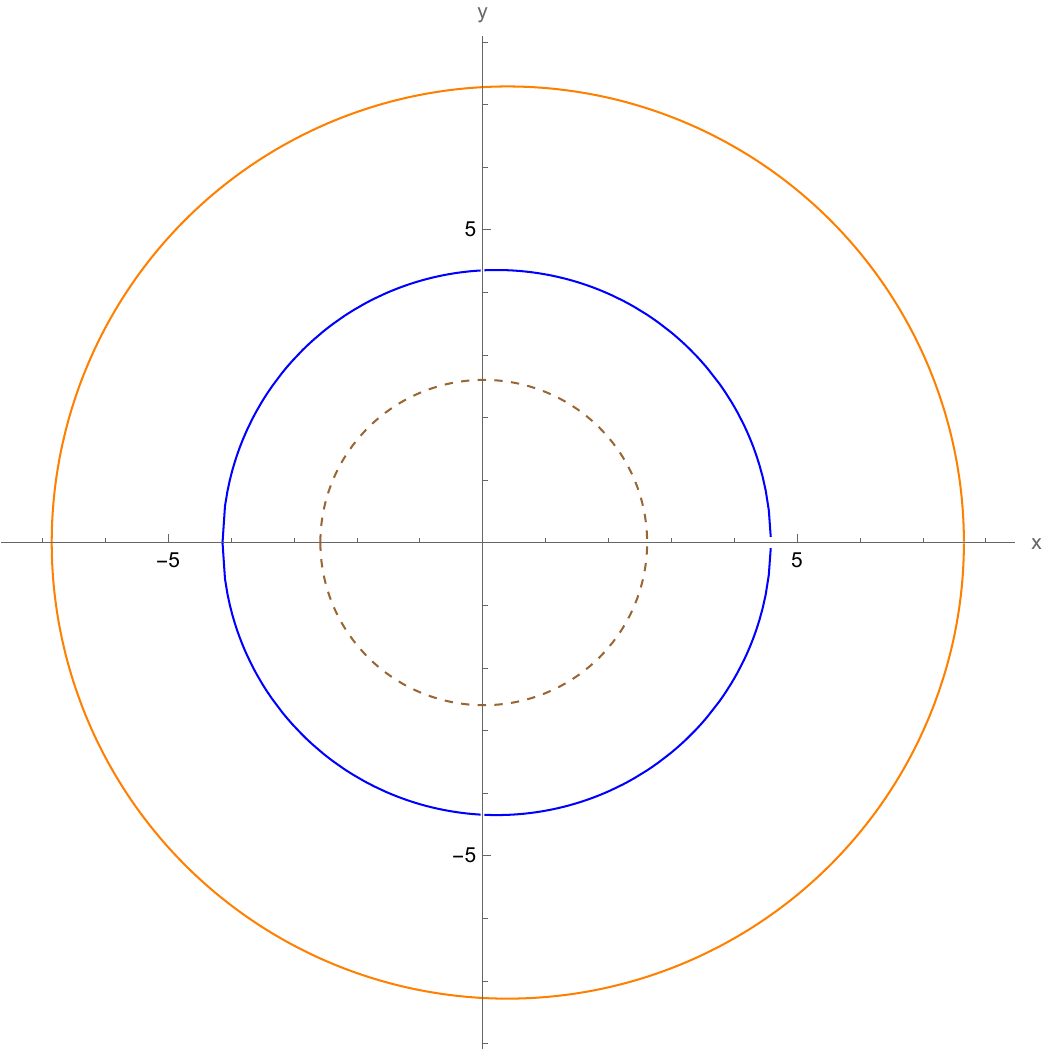}
        \par\vspace{1mm}
        \small{(a)}
		\end{minipage}%
		\begin{minipage}[t]{0.45\linewidth}
			\centering
			\includegraphics[height=6cm]{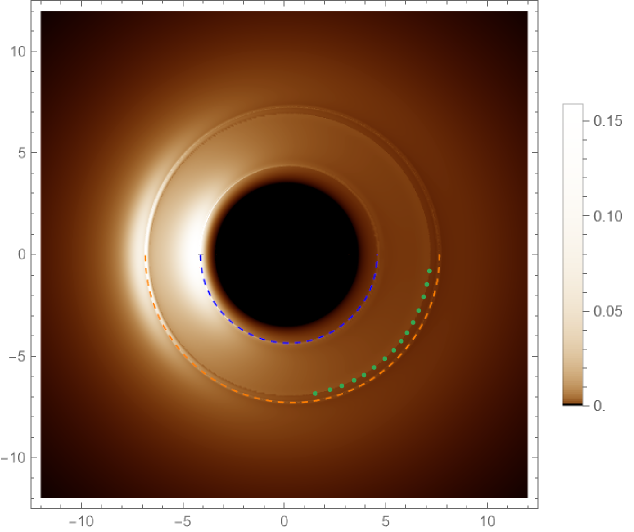}
        \par\vspace{1mm}
        \small{(b)}
		\end{minipage}%

		\begin{minipage}[t]{0.45\linewidth}
			\centering
			\includegraphics[height=6cm]{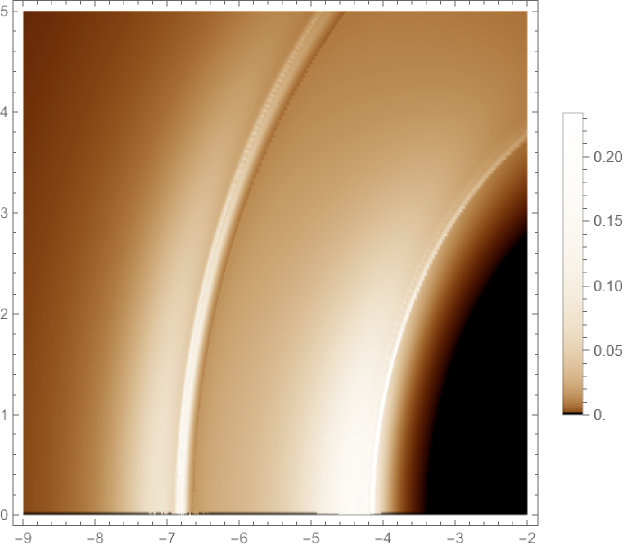}
        \par\vspace{1mm}
        \small{(c)}
		\end{minipage}%
		\begin{minipage}[t]{0.45\linewidth}
			\centering
			\includegraphics[height=6cm]{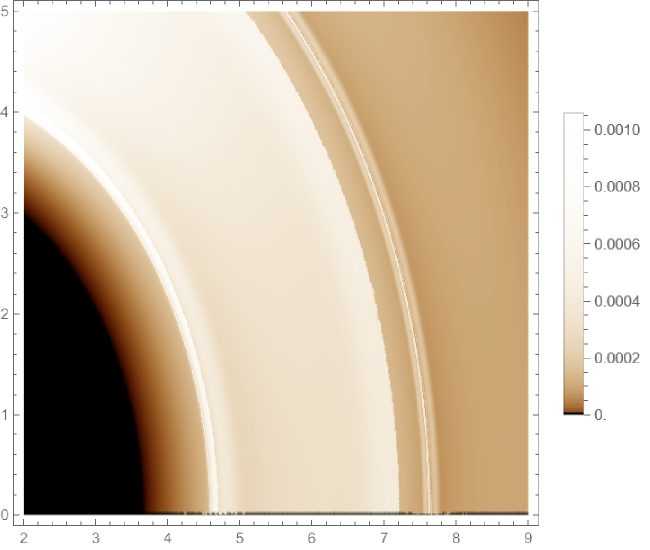}
        \par\vspace{1mm}
        \small{(d)}
		\end{minipage}%

  \caption{The shadow (a), images (b) and enlarged view of the image (c) and (d) of slowly rotating Kerr black hole with $m_-=0.5,a_-=0.01,m_+=1.4,a_+=0.2,R=3.8$.
  In (a), the orange curve represents the shadow boundary corresponds to the outer photon region and the blue one corresponds to the inner photon region.
   The brown dashed curve represents the shadow boundary corresponds to the inner photon region when no thin shell is present.
     In (b), the dashed orange and blue curve represents the outer and inner photon ring, respectively.
     The green dashed curves indicates the ``step" feature resulting from a discontinuity in the redshift factor.}
  \label{m0.5R3.8}
\end{figure}

The shadow and image are shown as Fig. \ref{m0.5R3.8}.
The photon regions are identified at $r_{PR-}\subseteq(1.49,1.51)$ for the interior and $r_{PR+}\subseteq(3.95,4.42)$ for the exterior, respectively. 
In this standard configuration, a double photon-ring structure indeed appears, corresponding to the two complete photon regions, as expected for spacetimes with two photon regions. In addition, the presence of the step-like structure indicates that it is a generic feature of the thin shell spacetime.

\subsection{$m_-=0.6,a_-=-0.3,m_+=1.13,a_+=0.1,R=2.7$.}

\begin{figure}[htbp]
		\begin{minipage}[t]{0.45\linewidth}
			\centering
			\includegraphics[height=6cm]{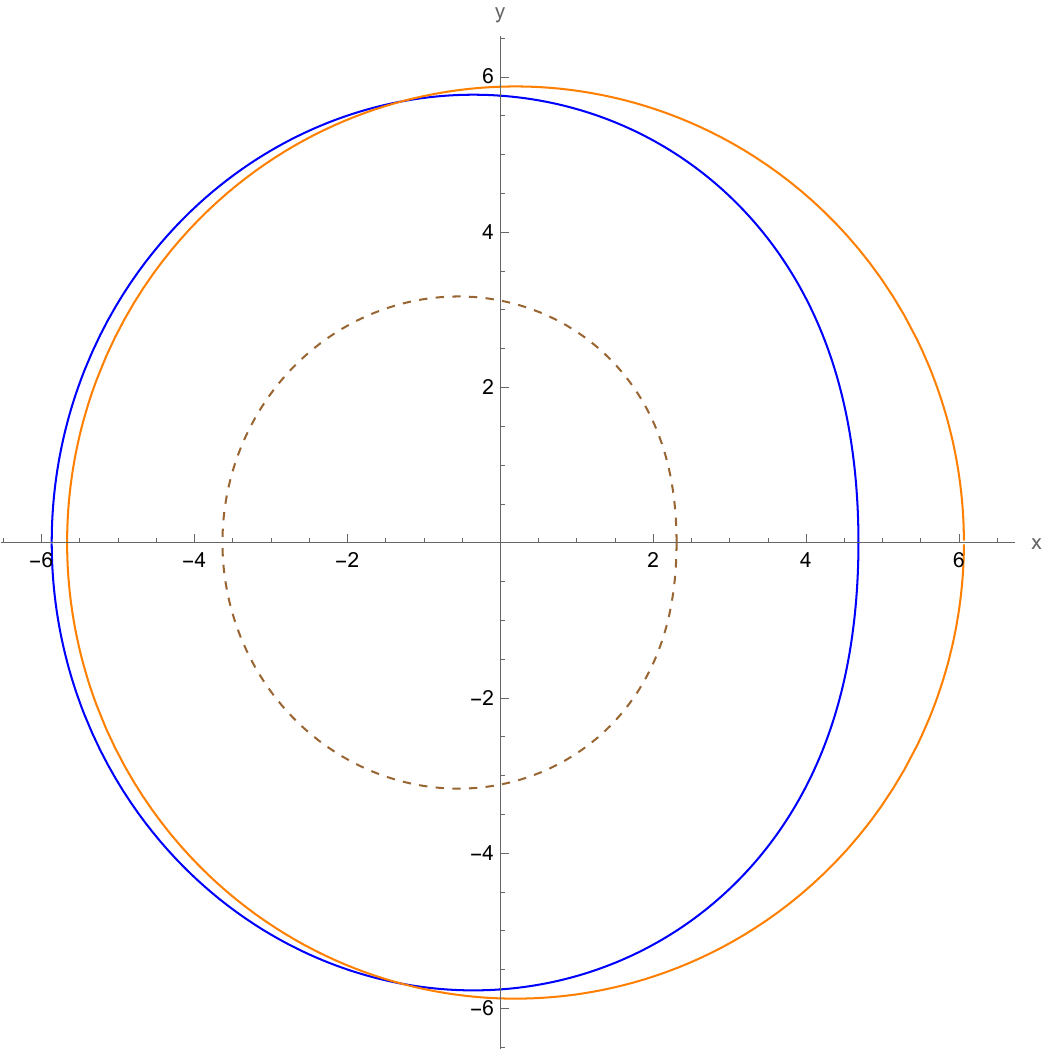}
        \par\vspace{1mm}
        \small{(a)}
		\end{minipage}%
		\begin{minipage}[t]{0.45\linewidth}
			\centering
			\includegraphics[height=6cm]{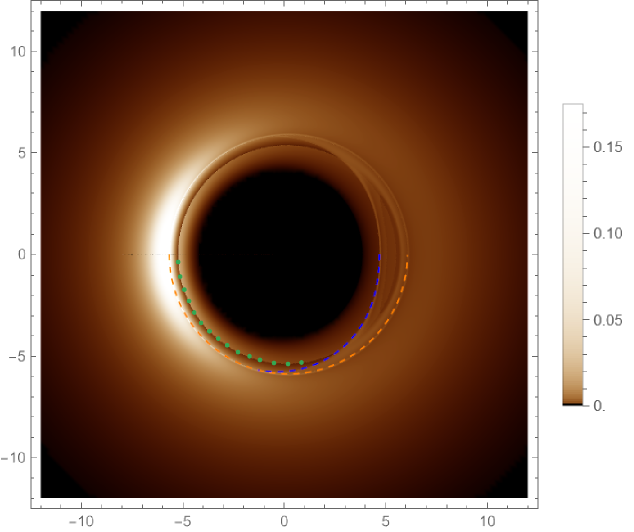}
        \par\vspace{1mm}
        \small{(b)}
		\end{minipage}%

		\begin{minipage}[t]{0.45\linewidth}
			\centering
			\includegraphics[height=6cm]{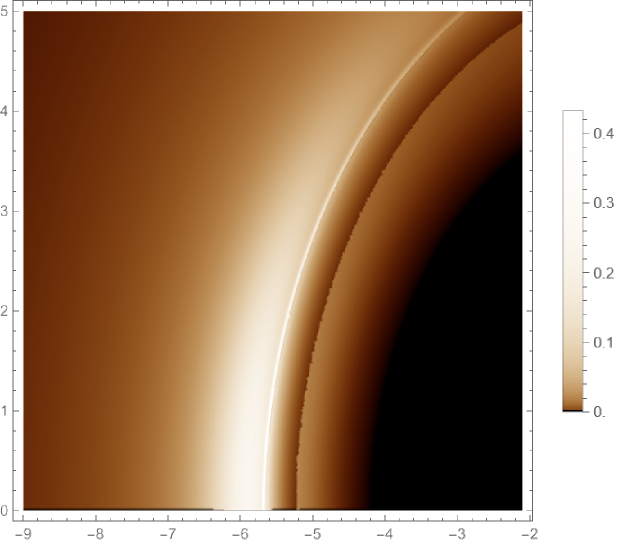}
        \par\vspace{1mm}
        \small{(c)}
		\end{minipage}%
		\begin{minipage}[t]{0.45\linewidth}
			\centering
			\includegraphics[height=6cm]{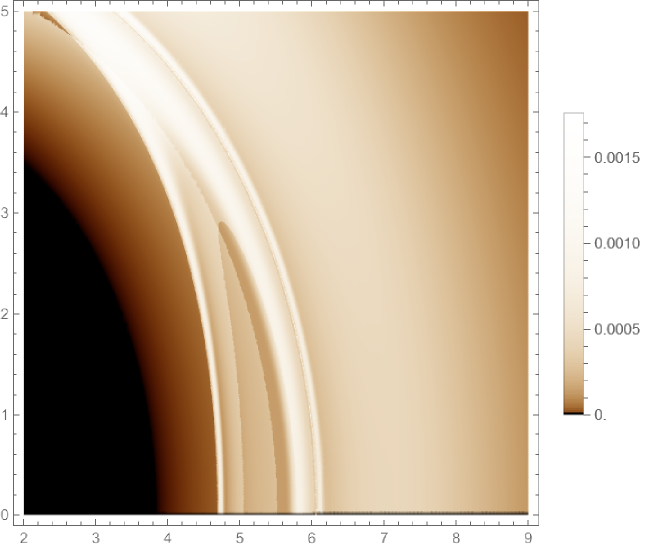}
        \par\vspace{1mm}
        \small{(d)}
		\end{minipage}%

  \caption{The shadow (a), images (b) and enlarged view of the image (c) and (d) of slowly rotating Kerr black hole with $m_-=0.6,a_-=-0.3,m_+=1.13,a_+=0.1,R=2.7$.
  In (a), the orange curve represents the shadow boundary corresponds to the outer photon region and the blue one corresponds to the inner photon region.
   The brown dashed curve represents the shadow boundary corresponds to the inner photon region when no thin shell is present.
     In (b), the dashed orange and blue curve represents the outer and inner photon ring, respectively.
     The green dashed curves indicates the ``step" feature resulting from a discontinuity in the redshift factor.}
  \label{m0.6a-}
\end{figure}

The shadow and image are shown as Fig. \ref{m0.6a-}.
The photon regions are identified at $r_{PR-}\subseteq(1.33,2.10)$ for the interior and $r_{PR+}\subseteq(3.27,3.50)$ for the exterior, respectively.
Under these circumstances, the direction of rotation is opposite between the interior spacetime and the exterior spacetime. 
As a result, the shadow boundary of the interior spacetime takes on an inverted D-shape (as shown by the blue curve in  Fig. \ref{m0.6a-} (a)). 
This image is similar to Fig. \ref{m0.6} (b), except that the double photon rings are located on the right side.

\section{Conclusion and discussion}\label{S4}

We study the image of the slowly rotating Kerr spacetime divided by a thin shell.
The shell is situated at radius $R$ , with zero angular momentum.
The interior spacetime is characterized by mass and spin parameters $m_-$ and $a_-$ , whereas the exterior spacetime has parameters $m_+$ and $a_+$, and $(a_-/m_-)^2, (a_+/m_+)^2 \ll 1$.
Using the junction conditions and neglecting $\mathcal{O}(a^2)$ terms, we find that when rays cross the shell, their energy $E$ changes, while the angular momentum $L$ and Carter constant $C$ remain conserved. 
This results in a transformation of the impact parameters $\eta$ and $\xi$, as given by \eqref{eta} and \eqref{xi}.

Using this transformation, we investigate the shadow and imaging features of this spacetime. 
Due to the presence of the shell, the redshift factor undergoes a discontinuous jump as the intersection of rays and accretion disk cross the shell. 
Moreover, the inner and outer photon regions may be truncated by the shell, resulting in incomplete photon region and a correspondingly non-closed shadow boundary. 
We represent such truncated shadows with dashed curves.
These truncated photon regions and shadows significantly influence the observed image, giving rise to novel and distinctive features. 
We clarify how the shell modifies the relation between shadow boundaries and the appearance of photon rings in the image, 
which exhibit the following characteristics:

\begin{itemize}
  \item The double photon ring structure emerges when the shadow of the outer photon region (outer shadow) overlaps with that of the inner photon region (inner shadow).
  
  \item In cases where the inner shadow extends beyond the outer shadow, no corresponding inner photon ring is visible on the image. 
  This phenomenon occurs because the effective potential corresponding to the inner photon region is lower than that of the outer region, rendering the inner photon ring invisible. 
  As seen in Fig. \ref{m0.6} (b), clear double photon rings are present on the left side; these rings gradually merge into a single ring toward the right, and ultimately only the outer photon ring remains.
  
  \item When the outer photon region is truncated by the shell, the outer photon ring may still appears in the image—even though rays do not actually reach the truncated portion of the photon region. 
  The shadow of the truncated region is indicated by a dashed curve. 
  Rays with impact parameters near these dashed curves orbit the black hole multiple times before reaching the shell, intersecting the accretion disk numerous times along the way. 
  Each intersection contributes to the observed intensity, resulting in the appearance of a photon ring or lensed ring in the image—even though the corresponding parts of the photon regions are theoretically cut off. 
  This, along with the previous one, indicates that the shadow and the photon ring no longer maintain a one-to-one correspondence.
  A similar phenomenon has been observed in spherically symmetric cases.
  
  \item A bright spot appears near the photon ring on the left side of the image. 
   This is caused by a discontinuous change in the redshift factor when the intersection point of the light ray with the accretion disk passes through the shell.
  
  \item A step-like feature appears in the image. 
  Similar to the mechanism above, this is caused by a sudden change in the redshift factor when the intersection points of the rays with the accretion disk traverse the shell.
  This characteristic has also been noted in scenarios of spherical symmetry.
  
  \item The V-shaped profile of the transfer function also occurs, analogous to that of a Schwarzschild spacetime with a shell.
  
\end{itemize}

Indeed, in modified gravity theories the Israel junction conditions are altered, and it remains unclear whether the photon four-momentum remains continuous across the shell.
Nevertheless, if the transformation law for the photon four-momentum across the shell can be derived from the junction conditions of modified gravity, our method can still be applied to compute the changes in the relevant conserved quantities, and subsequently to study the black hole image in such spacetimes.


Our future work will focus on Kerr spacetime with a shell, extending the analysis beyond the slow-rotation approximation by fully accounting for arbitrary spin $a$, and will explore alternative junction conditions.
Additionally, the case of Reissner-Nordström spacetime with a shell will be examined.
Furthermore, by applying the method of \cite{Azreg-Ainou:2014pra} to generalize a static wormhole to axisymmetry and employing the approach presented in this paper, we will be able to obtain an intuitive image of the slowly rotating wormhole.
Since the Israel junction condition is widely applied across various physical domains, our studies on black hole images with a thin shell will provide valuable insights. 
Ultimately, this research aims to assess, through future astronomical observations, the suitability of the Israel junction condition in describing real astrophysical systems.

\section*{Acknowledgement}
This research is supported in part by the National Natural Science Foundation of China under Grant Nos. 12588101, 12475063 and 12535002.

\end{document}